\documentclass[seceq]{ptptex}

\usepackage[usenames]{color}
\usepackage{graphicx}

\title{
Accurate Modelling of Relativistic Iron Lines from Accretion Discs%
}

\author{
Kris {\sc Beckwith} and Chris {\sc Done}%
}

\inst{
Department of Physics, University of Durham, South Road, Durham DH1 3LE, UK
}

\abst{
Observations of fluorescent iron lines from accreting black holes
provide one of the best tests of strong field
gravity available to date, and the only current observational tool to probe
black hole spacetime. 
However, the two most widely used models for spectral fitting
(diskline, laor) are over a decade old and have significant limitations.
We present a new code for calculating these effects which will be incorporated within 
the XSPEC package.
}

\begin{document}

\maketitle

\section{Introduction}

Material in an accretion disk around a black hole is orbiting at high
velocity, close to the speed of light, in a strong gravitational
potential. Hence its emission is distorted by doppler shifts, length
contraction, time dilation, gravitational redshift and
lightbending. The combined impact of these special and general
relativistic effects was first calculated in the now seminal paper of
Cunningham (1975),\cite{rf:1} \ where he used a \emph{transfer
function} to describe the relativistic effects. Any sharp spectral
features, such as the iron fluorescence line produced by X-ray
illumination of an accretion disc is transformed into broad, skewed
profile whose shape is given {\em directly} by the transfer function.\cite{rf:2} \ 

Observationally, evidence for a relativistically smeared iron line
first came from the ASCA observation of the active galactic nuclei
(AGN) MCG-6-30-15.\cite{rf:3} \ Further observations showed
evidence for the line profile being so broad as to require a maximally
spinning black hole.\cite{rf:4} \ More recent data from
XMM are interpreted as showing that the line is even wider than 
expected from an extreme Kerr disk, requiring direct 
extraction of the spin energy from the central black hole 
as well as the immense gravitational potential.\cite{rf:5}

While there are many caveats on extracting the line profile from the continuum,\cite{rf:7}
the dramatic results from MCG-6-30-15 plainly also require
that the extreme relativistic effects are well modelled.  There are
two models which are currently widely available to the observational
community, within the \texttt{XSPEC} spectral fitting package, \texttt{diskline} \cite{rf:2} and \texttt{laor}. \cite{rf:6} \ The analytic
\texttt{diskline} code models the line profile from an accretion disc
around a Schwarzschild black hole (so of course cannot be used to
describe the effects in a Kerr geometry). Also, it does not include
the effects of lightbending and hence does not accurately calculate all the
relativistic effects for $r< 20r_{g}$ (where $r_g=GM/c^2$).  
By contrast, the \texttt{laor} model numerically calculates the line profile 
including lightbending for an extreme Kerr black hole, but uses a rather 
small set of tabulated transfer functions which limit its resolution and accuracy
(see Section 3).  

While there are other relativistic codes in the literature which do
not suffer from these limitations, these are not generally readily
and/or easily available for observers to use. There is a clear need
for a fast, accurate, high resolution code which can be used to fit
data from the next generation of satellites. Here we give brief
results from our new code for computing the relativistic iron line
profile in both Schwarzschild and Kerr metrics. We compare this with
the \texttt{diskline} and
\texttt{laor} models in \texttt{XSPEC}, demonstrating the limitations
and assumptions inherent in these previous codes.

\section{Calculating Strong Gravitational Effects}

A line of rest energy $E_{int}$ emitted with rest frame emissivity on
the disc which we assume can be separated into 
radial ($r_e$) and angular $\cos\theta_e=\mu_e$ components through
$\epsilon(r_e,\mu_e)=\varepsilon(r_e)f(\mu_e)$. The flux as 
measured by a distant observer is\cite{rf:1}
\begin{align}
  \label{eqn:2.1.3}
    F_{o} \left( E_{o} \right) =  \frac{1}{r_{o}^{2}} \int \int  g^{4}
    \epsilon \left( r_{e},\mu_{e} \right) \delta \left( E_{o} - gE_{int} \right) 
d\alpha d\beta
\end{align}
where $g=E_o/E_{int}$ is the redshift factor and 
$d\alpha d\beta$ is the solid angle subtended by each small
patch of the disc in the observers frame of reference. If there is no
lightbending then this solid angle can be calculated analytically
by a transformation of variables to $dr_e dg$.\cite{rf:2}
This is the approach taken by the {\tt diskline} code\cite{rf:2}
making it very fast, but with obvious limitations for situations
where there is strong lightbending. 

Including lightbending means that the solid angle is much more complex
to calculate as a range of $\mu_{e}$ can contribute to a given
observed inclination angle, and these $\mu_e$ can only be found by
determining the full general relativistic light travel paths which
link the disc to the observer. Several attempts to include this in an
analytic transformation $d\alpha d\beta\to dr_e dg$
exist in the literature, but none of them are formally correct.\cite{rf:7}
A numerical finite differences approach has subtle numerical
pitfalls,\cite{rf:7} while a simple Monte-Carlo technique 
suffers from resolution problems due to the finite number of photons
which can be followed, and by the width of each 
radial ring on the disc and and the angular size of the observers
bin. Instead we use a geometric technique, based solely on the image
of the accretion disc at the observer. We use the analytic solutions
to find all the light travel paths,\cite{rf:7} which connect the disc
to the observer, and sort these paths by redshift factor $g$. We
use adaptive grids on  the image to find the boundaries of constant
$g$, and calculate the area $d\alpha d\beta$ directly from this.\cite{rf:7}

\section{Relativistic Line Profiles}

We have taken a disc from $r_{min} = 6r_{g}$ (the
minimum stable orbit for the Schwarzschild solution) to $r_{max} = 20
r_{g}$ (beyond which strong gravitational effects become of
diminishing importance) for both the Schwarzschild ($a = 0$) and
maximal Kerr ($a = 0.998$) cases for $\theta_{o} = 30^{\circ}$ and
$\varepsilon(r_e)\propto r_e^{-3}$. 

\begin{figure}
  \leavevmode
  \begin{center}
  \begin{tabular}{cc}
  \includegraphics[width=0.45\textwidth]{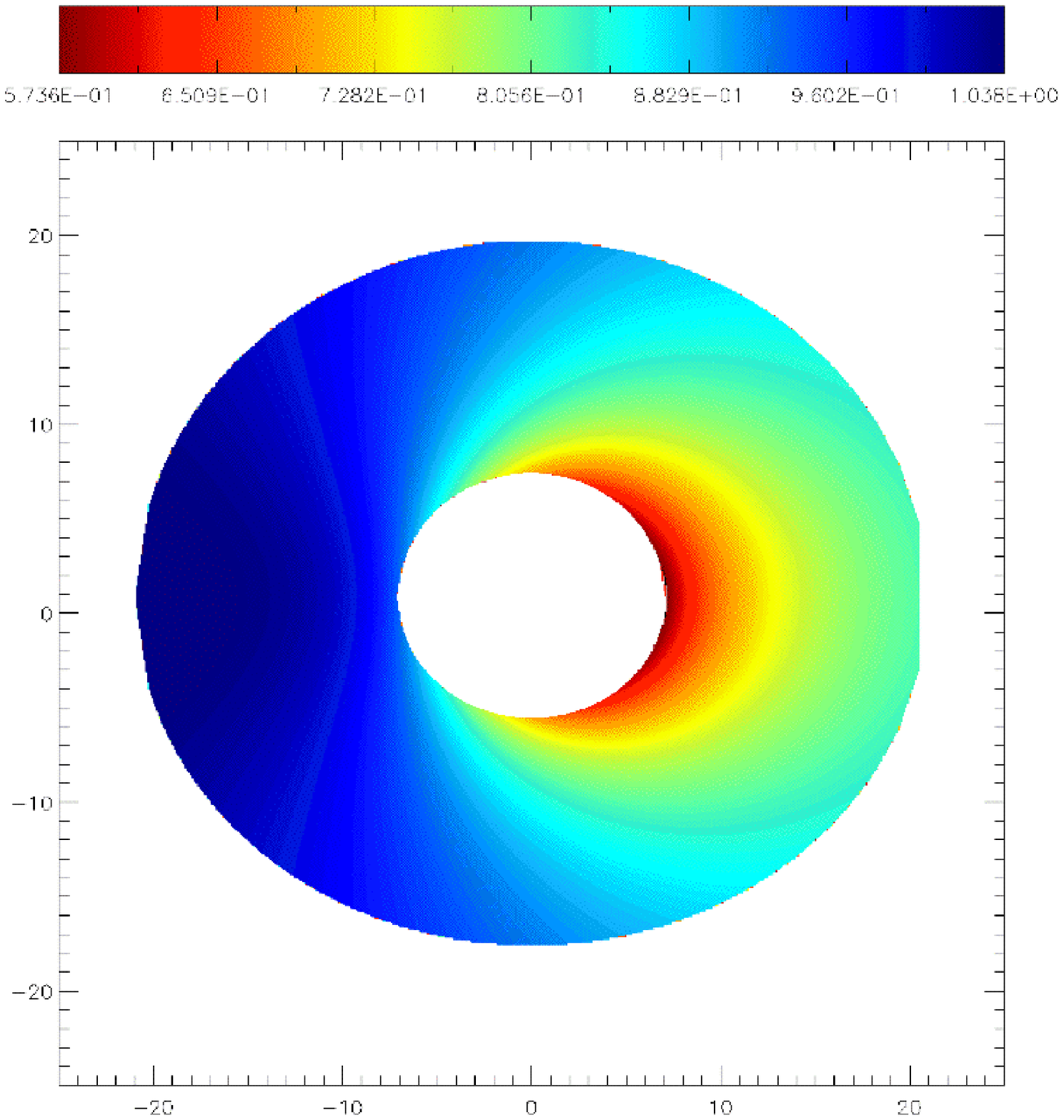} &
  \includegraphics[width=0.45\textwidth]{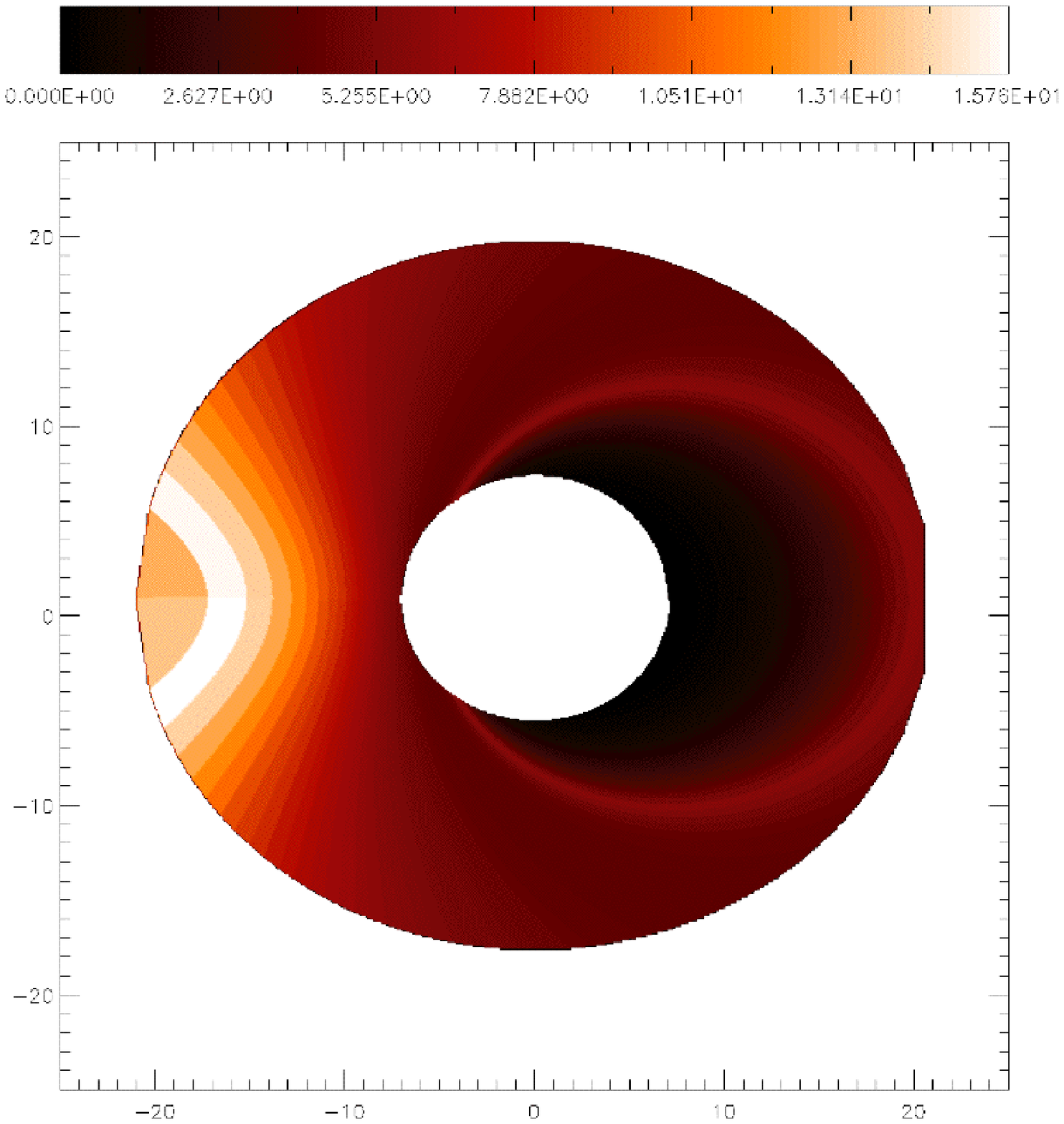}
  \end{tabular}
  \caption{Redshift images (left panel) and flux images (right panel) of the accretion disc on the $\left(\alpha, \beta \right)$ plane for a Schwarzschild black hole and observer with inclination of $30^{\circ}$}
  \label{fig:3.1.1}
  \end{center}
\end{figure}

\begin{figure}
  \leavevmode
  \begin{center}
  \begin{tabular}{ccc}
  \includegraphics[width=0.3\textwidth]{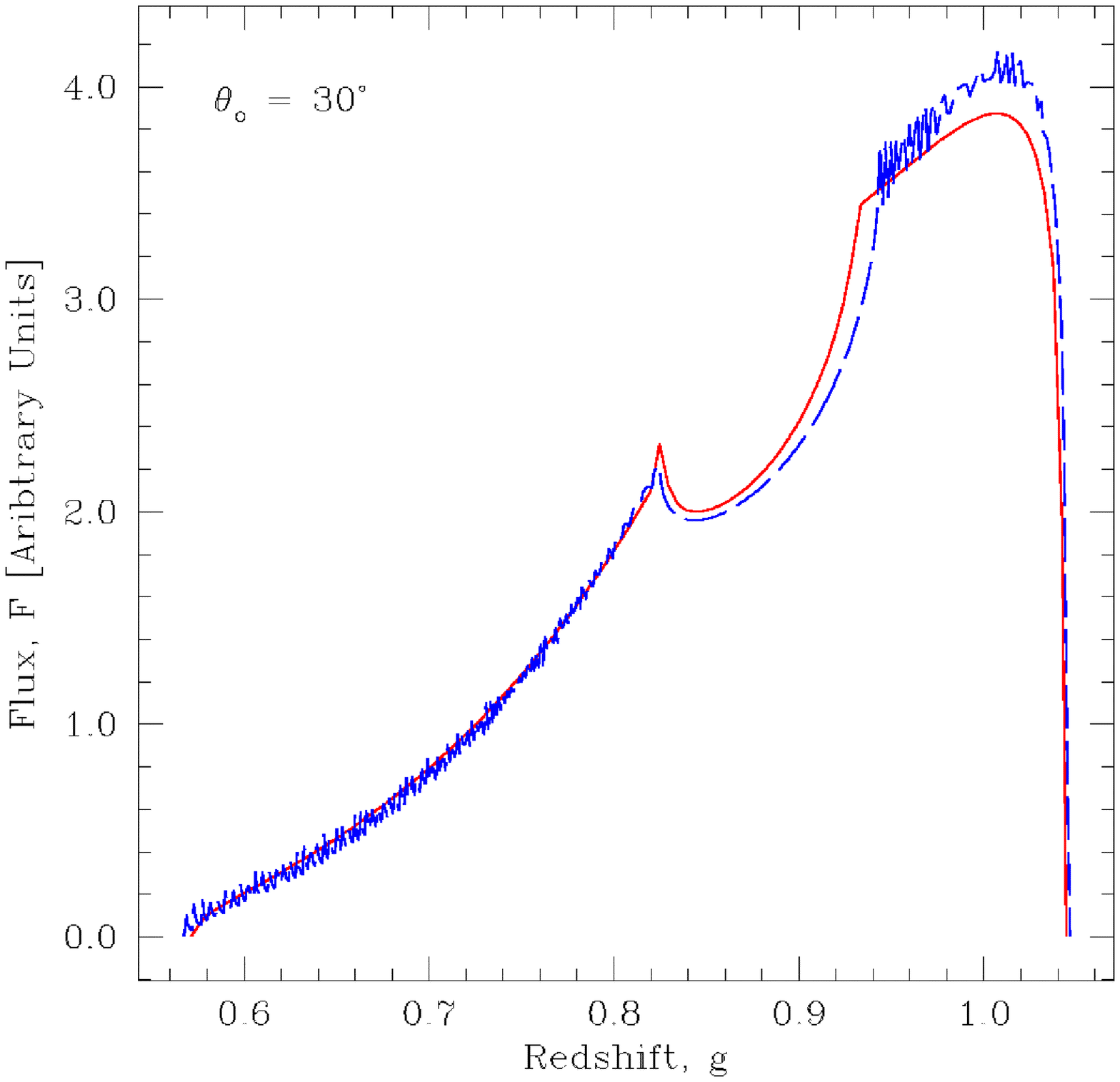} &
  \includegraphics[width=0.3\textwidth]{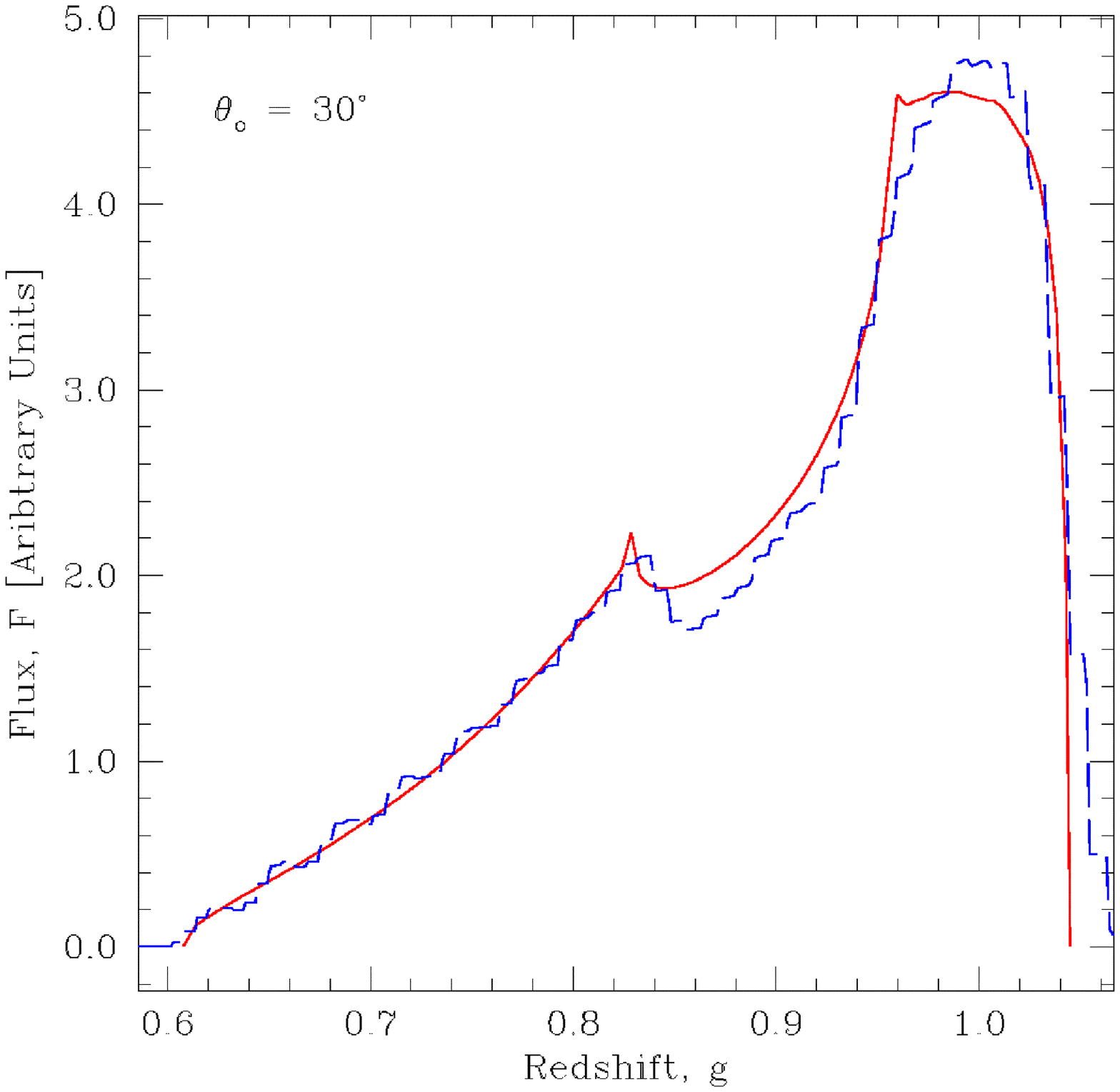} &
  \includegraphics[width=0.3\textwidth]{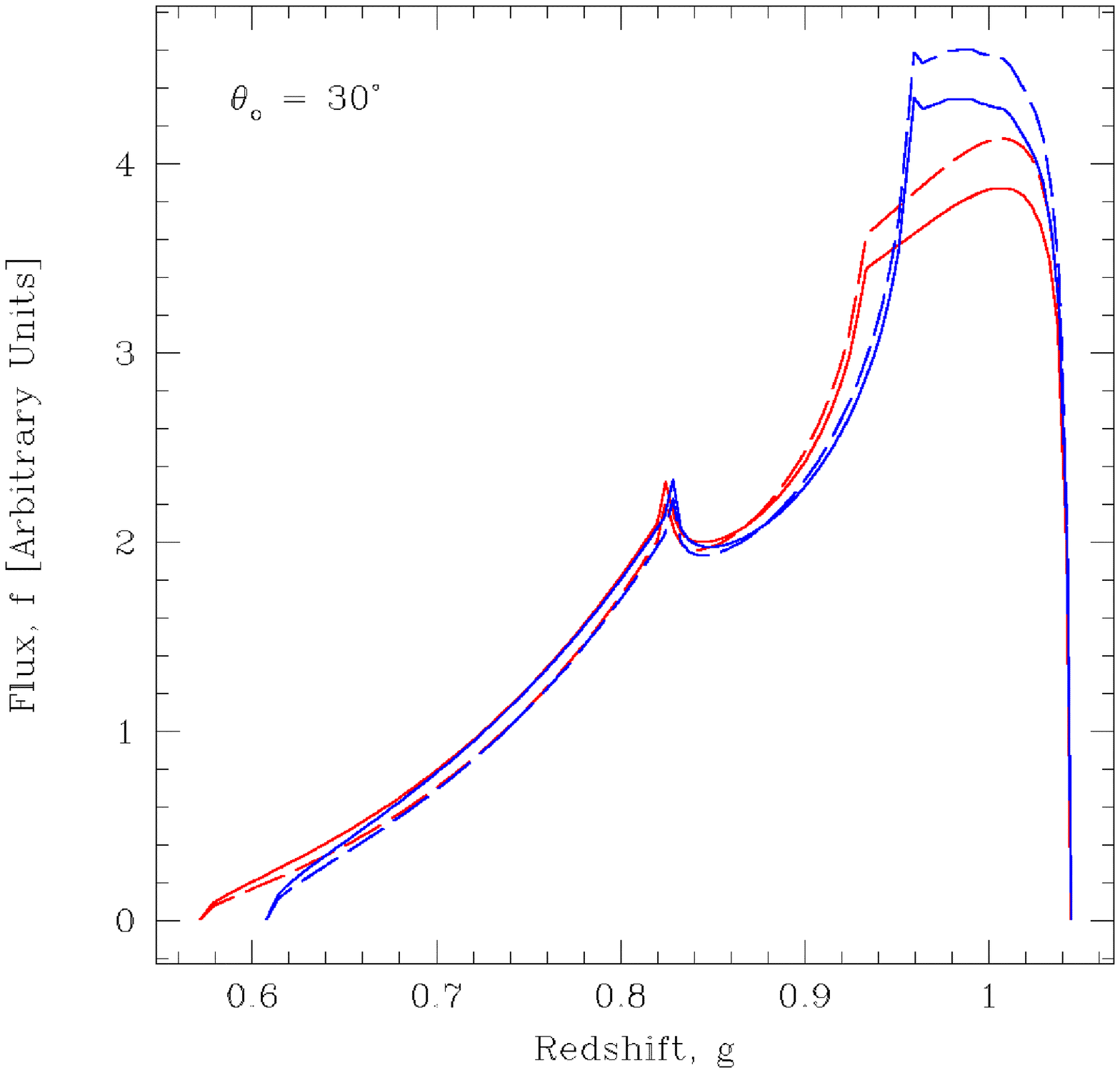}
  \end{tabular}
  \caption{Comparison of the relativistic line profile computed by our model 
({\color{red} red} solid line) with that computed by the \texttt{XSPEC diskline} 
model ({\color{blue} blue} dashed line) for $\varepsilon \left( r_{e} \right) \propto r^{-3}$ and $f \left( \mu_{e} \right) = 1$ (left panel) and that computed by the \texttt{XSPEC laor} model 
({\color{blue} blue} dashed line) for $\varepsilon \left( r_{e} \right) \propto r^{-3}$ and $f \left( \mu_{e} \right) \propto \left(1+2.06\mu_e \right)$ (middle panel).  The right panel shows a comparison of the relativistic line profiles generated by our model with (a) $\varepsilon \left( r_{e} \right) \propto r_{e}^{-3}$, $f \left( \mu_{e} \right) = 1$ (solid lines) and (b) $\varepsilon \left( r_{e} \right) \propto r_{e}^{-3}$, $f \left( \mu_{e} \right)  \propto \left( 1 + 2.06 \mu_{e} \right)$ (dashed lines) for the case of Schwarzschild ({\color{red} red} lines) and maximal Kerr ({\color{blue} blue} lines) black holes.}
  \label{fig:3.1.2}
  \end{center}
\end{figure}

The {\tt diskline} code assumes a Schwarzschild metric ($a=0$) and
additionally that light travels in straight lines (so the angular
emissivity term is irrelevant). Hence we use $f(\mu_e)=1$ 
(no angular dependance of the emissivity). 
Figure \ref{fig:3.1.1} (left panel) shows our
redshift image of the Schwarzschild disc, where the colours
indicate the value of the redshift factor, $g$, whilst Figure
\ref{fig:3.1.1} (right panel) shows the corresponding flux image with each redshift
bin coloured by its area on the observers sky. 
Figure \ref{fig:3.1.2} (left panel) shows our line profile compared to 
that from the {\tt diskline} code. We see that our new model matches
very closely to the \texttt{XSPEC diskline} model for a nearly face on
disk. Whilst the key difference between our model and
\texttt{diskline} is the inclusion of light-bending effects, this has little effect
if there is no angular dependance to the emissivity.

By contrast, the \texttt{laor} code is based on transfer functions
calculated by Monte-Carlo methods over a range of radii in extreme Kerr, and
includes a standard limb darkening law $f(\mu_e)\propto
(1+2.06\mu_e)$.  We include this limb darkening in  our code, and Figure
\ref{fig:3.1.2} (middle panel) shows the line profile comparison. It is clear that there are some
resolution issues in the \texttt{XSPEC laor} model. 

The effects of spacetime and emissivity are shown in Fig. \ref{fig:3.1.2} (right panel)
with radial emissivity of $r^{-3}$ over 6--20 $r_g$.
The lines are (from top to bottom at the
blue peak) from extreme Kerr with limb darkening, extreme Kerr with
constant angular emissivity, Schwarzschild with limb darkening and
Schwarzschild with constant emissivity. There is a 40\% change in
relative height in the blue peak between the various models. In data
fitting, if the model assumed a Kerr metric with limb darkening
(i.e. {\tt laor}) while the real line was from a Schwarzschild,
constant angular emissivity disc, then a $\chi^2$ minimisation would
tend to match up the blue peak heights. This would result in a deficit
across the red wing, pulling the {\em radial} emissivity into a more
centrally peaked value for $q$. We caution that there are significant
uncertainties in the angular distribution of the line emissivity which
can change the expected line profile due to lightbending even at
low/moderate inclinations.

\section{Conclusions}

We show results from our code for a disc between 6--20 $r_g$
with radial emissivity $\propto r^{-3}$ in both Schwarzschild and Kerr
metrics, comparing these with the {\tt diskline} and {\tt laor} models
in {\tt XSPEC}. Lightbending is {\em always} important, in that the
image of the disc at the observer {\em always} consists of a range of
different emission angles. This means that the angular dependance of
the emitted flux can make significant changes to the derived line profile. 
While calculating the strong gravity effects is a difficult numerical
problem, the underlying physics is well known. By contrast, the {\em
angular} emissivity is an astrophysical problem, and is not at all
well known as it depends on the ionisation state of the disc as a
function both of height and radius. Figure \ref{fig:3.1.2} (right panel) demonstrates the
effect of this unknown astrophysics folded through both Schwarzschild
and extreme Kerr spacetimes, showing there can be a 20\% difference in
the ratio of the blue-to-red peak heights simply from the assumed
angular emissivity. Before we can use the line profiles to
test General Relativity, to probe the underlying physics, we will
need to have a much better understanding of the astrophysics of
accretion.


%

\end{document}